\documentclass[aps,pra,twocolumn,showpacs,nofootinbib,amsmath]{revtex4}% superscriptaddress
\usepackage{epsfig}

\usepackage{amssymb}
\usepackage{bm}
\usepackage{epsf}

\DeclareFontFamily{OT1}{rsfs}{}
\DeclareFontShape{OT1}{rsfs}{m}{n}{<5> rsfs5 <7> rsfs7 <10>
rsfs10}{}
\DeclareSymbolFont{mathrsfs}{OT1}{rsfs}{m}{n}
\DeclareSymbolFontAlphabet{\mathrsfs}{mathrsfs}

\newcommand{\J}{\mathrsfs{J}}
\newcommand{\N}{\mathrsfs{N}}

\newcommand{\M}{\mathfrak{M}}

\newcommand{\A}{\mathbf{A}}
\newcommand{\e}{\mathbf{e}}
\newcommand{\E}{\mathbf{E}}
\newcommand{\K}{\mathbf{k}}
\renewcommand{\P}{\mathbf{p}}

\newcommand{\R}{\mathbf{r}}

\newcommand*{\ket}[1]{\left| #1 \right\rangle}
\newcommand*{\bra}[1]{\left\langle{#1}\right|}

\newcommand{\SFA}{{\rm sfa}}
\newcommand{\SCV}{{\rm scv}}
\begin{document}

%\title{On Deficiencies of the Coulomb-Volkov Approximation for Description of Angle-Resolved Photoelectron Spectra in Strong-field ionization by a High Frequency Field}
\title{Photoelectron spectra in strong-field ionization by a high frequency field}

\author{Denys I. Bondar}
\email[To whom correspondence should be addressed:\ ]{dbondar@sciborg.uwaterloo.ca}
\affiliation{University of Waterloo, Waterloo, Ontario N2L 3G1, Canada}
\affiliation{National Research Council of Canada, Ottawa, Ontario K1A 0R6, Canada}

\author{Michael Spanner}
\affiliation{National Research Council of Canada, Ottawa, Ontario K1A 0R6, Canada}

\author{Wing-Ki Liu}
\affiliation{University of Waterloo, Waterloo, Ontario N2L 3G1, Canada}

\author{Gennady L. Yudin}
\affiliation{Universit\'{e} de Sherbrooke, Sherbrooke, Qu\'{e}bec J1K 2R1, Canada}
\affiliation{National Research Council of Canada, Ottawa, Ontario K1A 0R6, Canada}

\date{\today}

\begin{abstract}
We analyze atomic photoelectron momentum distributions induced by bichromatic and monochromatic laser fields within the strong field approximation (SFA), separable Coulomb-Volkov approximation (SCVA), and {\it ab initio} treatment. 
We focus on the high frequency regime -- the smallest frequency used is larger than the ionization
potential of the atom. We observe a remarkable agreement between the {\it ab initio} and velocity gauge SFA
results while the velocity gauge SCVA fails to agree. Reasons of such a failure are discussed. 
\end{abstract}

\maketitle

\section{Introduction}

The simultaneous application of more than one laser field leads to a number of novel effects. These effects
include reducing the excitation or ionization rates \cite{Chen1990} or altering the photoelectron angular
distributions and the harmonic emission parity selection rules \cite{Potvliege1992a}.  Furthermore, a laser
field can dress or strongly mix the field-free excited states, including the continuum, of an atom. This
produces new resonance-like structures, which could not exist otherwise.  Such an effect is called
laser-induced continuum structure \cite{Faucher1993, Knight1990,  Bondar1995}, and it has been applied to
design lasers without inversion \cite{Imamolu1991}.  Multicolor lasers are also exploited for control of
quantum dynamics \cite{Shapiro1997, Shapiro2006, Shapiro2003}, laser-induced electron diffraction
\cite{Murray2008}, attosecond laser pulse synthesis \cite{Fleischer2006}, and above-threshold ionization (ATI) 
\cite{Bucksbaum1990,Bucksbaum1994,Meyer2008}.  Moreover, multistep ionization, 
where each step is driven by a laser at its resonant
frequency, has many useful applications: efficient atomic isotope separation \cite{Shore1990}, extremely
sensitive detection of small numbers of atoms in a sample (single-atom detection) \cite{Hurst1979}, and
very-high-resolution photoelectron spectroscopy \cite{Kimura2007}.  Nevertheless, the presented list of
phenomena is not exhausted, and we refer the reader to reviews \cite{Astapenko2006, Potvliege1992a,
Ehlotzky2001} for further discussions.

Perelomov and Popov were the first to study theoretically ionization of an atom by multi-color laser radiation
\cite{Perelomov1967a} within the imaginary-time method \cite{Popov2005}.  Afterwards, a broad variety of
perturbative \cite{Pazdzersky1997, Fifirig1997, Fifirig2001, Fifirig2004} as well as non-perturbative results
\cite{Zhou1991, Baranova1993, Pazdzersky1995, Kuchiev1998, Kuchiev1999, Pazdzersky2005, Koval2007} has been
obtained. In addition, extensive numerical investigations have also been performed \cite{Potvliege1991, Chu1991, Dorr1991, Schafer1992, Potvliege1992, Potvliege1994, Protopapas1995,
Chu2004}.

In the present paper, we first present a general formalism to tackle the problem of ionization by multiple laser
fields with no restriction on the parameters of the laser fields. Two analytical approaches are employed: the
strong field approximation (SFA) and the separable Coulomb-Volkov approximation (SCVA). However, as it will be
seen below, both the approximations lead to quite different results for photoelectron momentum distributions.  Therefore, {\it ab initio} calculations
have been done in order to be able to make conclusions upon the validity of our analytical predictions.

The main result of this paper is that on the one hand, the SCVA fails to  agree with {\it ab initio} results; on the other hand,  the SFA adequately matches with {\it ab initio} calculations. We will discuss causes of such a fact.

The SFA and SCVA are diametrical opposites in treating the final state of a photoelectron.  The former
approximation completely ignores the influence of the Coulomb field of an ion by employing the Volkov wave
function for the final state of an electron \cite{Keldysh1965a, Faisal1973, Reiss1980a}  (atomic units are used throughout), 
\begin{equation}
\ket{\Psi_{\SFA}(t)} = \exp\left\{ -\frac i2 \int^{t} [\P + \A(\tau)]^2 d\tau \right\} \ket{\P},
\end{equation}
where $\A(t)$ is the vector potential of a laser field and $\ket{\P}$ denotes a plane wave with
momentum $\P$.  On the contrary, the
SCVA attempts to account for the Coulomb correction by using the separable Coulomb-Volkov continuum wave
function (or the Coulomb-Volkov ansatz) that is defined as 
\begin{equation}\label{SCV}
\ket{\Psi_{\SCV}(t)} = \exp\left\{ -\frac i2 \int^{t} [\P + \A(\tau)]^2 d\tau \right\} \ket{\psi_{\P}},
\end{equation}
where $\ket{\psi_{\P}}$ denotes a stationary laser-free
atomic continuum wave function with a defined asymptotic momentum $\P$ [the wave function $\ket{\psi_{\P}}$
satisfies the orthonormality condition $\bra{\psi_{\P}} \psi_{\P'} \rangle = \delta(\P-\P')$]. A justification
for the SCVA based on the sudden perturbation approximation \cite{Dykhne1978, Dykhne1977} and the sequential
three-stage model of laser-assisted x-ray photoionization has been presented in Refs. \cite{Yudin2007,
Yudin2008}, where the SCVA has been used to refine the theory of an attosecond streak camera and spectral phase
interferometry.

Recent developments and projects of generation of high-intensity x-ray laser sources (see, e.~g., Refs.
\cite{Makris2008,Moore2008,Johnsson2008,Sorokin2007,Ayvazyan2006,Tsakiris2006,Wabnitz2005} and references
therein) have stimulated interest in the theoretical description of multiphoton ionization at high laser
frequencies. Thus, we illustrate our results for the case when the smallest frequency of a laser is larger than
the ionization potential of an atom.

The rest of the paper is organized as follows. In Sec. \ref{Sec2}, the general formalism for obtaining the
velocity gauge SFA and SCVA multicolor ionization amplitudes is developed.  In Sec. \ref{Sec3}, ionization by
a bichromatic laser field is investigated by means of both the approximations as well as {\it ab initio}
numerical solution of the Schr\"{o}dinger equation. Ionization by  monochromatic laser radiation is studied in Sec. \ref{Sec4}. Finally, conclusions are made in the last section.

\section{Mathematical Background}\label{Sec2}

Before presenting our general scheme, we clarify the main peculiarity of our formalism.  The saddle point
approximation was applied to calculate photoionization amplitudes in the majority of previously published
approaches.  Rigorously speaking, the use of the saddle point approximation inevitably poses some restrictions
on the parameters of laser fields.  To avoid any restrictions, we generalize the ideology of the
Keldysh-Faisal-Reiss theory \cite{Keldysh1965a, Faisal1973, Reiss1980a} (and its recent applications to
ionization in bichromatic  laser radiation \cite{Koval2007, Pazdzersky2005}) and expand the time-dependent part
of the Volkov wave function [see Eq. (\ref{TimeDepVolkov})] into a multiple Fourier series.  Such an expansion
allows us to obtain readily close analytical expressions for the SFA and SCVA ionization amplitudes in the
velocity gauge.

The velocity gauge SCVA and SFA photoionization amplitudes are defined as
\begin{eqnarray}
\M_{\SCV}  = -i\int_{-\infty}^{\infty} \bra{\Psi_{\SCV}(t)} \hat{\P}\cdot\A(t)  \ket{g}e^{iI_p t} dt, \label{Deff_M_SCVA} \qquad\qquad \\
\M_{\SFA}  =  -i\int_{-\infty}^{\infty} \bra{\Psi_{\SFA}(t)} \hat{\P}\cdot\A(t) + \frac 12 \A^2(t) \ket{g}e^{iI_p t} dt. \label{Deff_M_SFA}
\end{eqnarray} 
Here $I_p$ is the ionization potential, $\ket{g}$ is the initial atomic state, and the vector potential reads
$\A(t) = \sum_{n=1}^N \A_n(t)$, with
\begin{eqnarray}\label{DeffAbyE}
\A_n (t) = - \frac{\E_1^{(n)}}{\Omega_n}\sin(\Omega_n t+\varphi_n) - \frac{\E_2^{(n)}}{\Omega_n}\cos(\Omega_n t + \varphi_n),
\end{eqnarray}
where $\E_{1}^{(n)}$ and $\E_{2}^{(n)}$ are mutually orthogonal electrical fields, $\Omega_n$ is the frequency ($\Omega_n\neq 0$), and $\varphi_n$ is the absolute phase. Equation (\ref{DeffAbyE}) can be alternatively written as
\begin{eqnarray}
\A_n (t) &=& A_0^{(n)} \left[ \e_1^{(n)}\sin\left(\Omega_n t+\varphi_n\right)\sin\phi_n   \right. \nonumber\\ 
&& \qquad \left. + \e_2^{(n)}\cos\left(\Omega_n t+\varphi_n\right)\cos\phi_n \right], \nonumber
\end{eqnarray}
where $A_0^{(n)}=- \mathrsfs{E}_n/\Omega_n$, $\mathrsfs{E}_n=\sqrt{\left(\E_1^{(n)}\right)^2 + \left(\E_2^{(n)}\right)^2}$, $\phi_n$ is the ellipticity parameter ($\sin\phi_n =E_1^{(n)}/ \mathrsfs{E}_n$, $\cos\phi_n =E_2^{(n)}/ \mathrsfs{E}_n$), and $\e_{1,2}^{(n)}$ are the polarization vectors along the vectors $\E_{1,2}^{(n)}$, respectively.

The time-dependent part of the Volkov wave function 
\begin{eqnarray}\label{TimeDepVolkov}
\Phi^{V} (t) = \exp\left\{ -\frac i2 \int^t \left[ \P + \sum_{n=1}^N \A_n(\tau) \right]^2d\tau \right\},
\end{eqnarray}
can be factorized in the following way: 
\begin{eqnarray}\label{VolkovFact}
\Phi^{V} (t) = \prod_{n=1}^N \left[ \Phi_n^{(1)} (t) \prod_{m=n+1}^N \Phi_{n,m}^{(2)}(t) \right],
\end{eqnarray}
where
\begin{eqnarray}
\Phi_n^{(1)} (t) &=& \exp\left\{  i\frac{N-1}{2N}\P^2 t-\frac i2 \int^t \left[ \P + \A_n(\tau) \right]^2d\tau \right\}, \nonumber\\
\Phi_{n,m}^{(2)}(t) &=& \exp\left\{ -i \int^t \A_n(\tau)\cdot\A_m(\tau) d\tau \right\}, \label{DeffPhi_One_Two}
\end{eqnarray}
The interpretation of factorization (\ref{VolkovFact}) is quite obvious: the term 
$\Phi_n^{(1)}(t)$ represents the contribution from propagation solely in the field 
$\A_n(t)$ and the term $\Phi_{n,m}^{(2)}(t)$ -- a coupling between propagations in 
the fields $\A_n(t)$ and $\A_m(t)$.
Carrying out tedious but straight forward calculations, we obtain the Fourier expansion 
of the components $\Phi^{(1)}$ and $\Phi^{(2)}$,
\begin{eqnarray}
\Phi_n^{(1)}(t) &=& \sum_{k=-\infty}^{\infty} i^k \exp\left\{ik\left(\varphi_n-\varphi_0^{(n)}\right)\right\} \nonumber\\
&& \times	\J_k \left( -\N_n, -\mathrsfs{M}_2^{(n)} ; 2\varphi_0^{(n)} \right) \nonumber\\
&& \times \exp\left\{ -i \left(\frac{\P^2}{2N} + \left[2\mathrsfs{M}_1^{(n)} - k\right]\Omega_n \right)t \right\}, \quad\label{FourierPhi1}\\ \Phi_{n,m}^{(2)}(t) &=& \sum_{k,l=-\infty}^{\infty} J_k (\mathrsfs{K}_{n,m}) J_l (\mathrsfs{L}_{n,m}) \nonumber\\
&&\times e^{i(l+k)\varphi_n + i(l-k)\varphi_m -ik\alpha_{n,m} -il\beta_{n,m}} \nonumber\\
&& \times \exp\left[ i(l+k)\Omega_n t + i(l-k)\Omega_m t \right], \label{FourierPhi2}
\end{eqnarray}
where
\begin{eqnarray}
&& \mathrsfs{M}_{1,2}^{(n)} = \frac 1{(2\Omega_n)^3}\left[ \left(\E_1^{(n)}\right)^2 \pm \left(\E_2^{(n)}\right)^2 \right], \nonumber\\
&& \N_n = \sqrt{ \left(\N_1^{(n)}\right)^2 + \left(\N_2^{(n)}\right)^2}, \quad
		\N_{1,2}^{(n)} = \frac{\P\cdot\E_{1,2}^{(n)}}{\Omega_n^2}, \nonumber\\
&& \mathrsfs{K}_{n,m} =-\sigma_1^{(n,m)} /[2\Omega_n\Omega_m(\Omega_n - \Omega_m)],  \nonumber\\
&& \mathrsfs{L}_{n,m} = -\sigma_2^{(n,m)}/[2\Omega_n\Omega_m(\Omega_n + \Omega_m)], \nonumber\\
&& \sigma_{1,2}^{(n,m)} = \left( \left[ \E_1^{(n)}\cdot\E_2^{(m)} \mp \E_2^{(n)}\cdot\E_1^{(m)}\right]^2  \right. \nonumber\\
&& \qquad\qquad \left. + \left[ \E_2^{(n)}\cdot\E_2^{(m)}\pm \E_1^{(n)}\cdot\E_1^{(m)} \right]^2\right)^{1/2}, \nonumber
\end{eqnarray}
The angle $\varphi_0^{(n)}$ is defined by the equations $\cos\varphi_0^{(n)} = \N_1^{(n)}/\N_n$ and $\sin\varphi_0^{(n)} = -\N_2^{(n)}/\N_n$; the angles $\alpha_{n,m}$ and $\beta_{n,m}$ are given as solutions of the following equations
\begin{eqnarray}
\cos\alpha_{n,m} &=& \left[ \E_1^{(n)}\cdot\E_1^{(m)} + \E_2^{(n)}\cdot\E_2^{(m)}\right]/\sigma_1^{(n,m)}, \nonumber\\
\sin\alpha_{n,m} &=& \left[ \E_1^{(n)}\cdot\E_2^{(m)} - \E_2^{(n)}\cdot\E_1^{(m)}\right]/\sigma_1^{(n,m)}; \nonumber\\
\cos\beta_{n,m} &=& \left[ \E_2^{(n)}\cdot\E_2^{(m)} - \E_1^{(n)}\cdot\E_1^{(m)}\right]/\sigma_2^{(n,m)}, \nonumber\\
\sin\beta_{n,m} &=& \left[ \E_1^{(n)}\cdot\E_2^{(m)} + \E_2^{(n)}\cdot\E_1^{(m)}\right]/\sigma_2^{(n,m)}. \nonumber
\end{eqnarray}
$\J_{n}\left(u,v;\eta\right)$ denotes a two-variable one-parameter Bessel function \cite{Korsch2006}
\begin{eqnarray}\label{2D-BessFunc}
\J_n \left(u,v;\eta \right) &=& \frac 1{2\pi} \int_{-\pi}^{+\pi} e^{i(u\sin t + v\sin(2t +\eta)-nt)} dt = \nonumber\\
&=& \sum_{k=-\infty}^{\infty} J_{n-2k}(u) J_k(v) e^{i k\eta},
\end{eqnarray}
where $J_n(x)$ is the well-known ordinary Bessel function.

It is convenient to write down separately the result for the case of multicolor 
linearly polarized fields ($\E_2^{(n)} = {\bf 0}$),
\begin{eqnarray}
&& \mathrsfs{N}_n = \mathrsfs{N}_1^{(n)}, \quad \mathrsfs{M}_1^{(n)} = \mathrsfs{M}_2^{(n)}= U_p^{(n)}/2\Omega_n, \nonumber\\
&& \varphi_0^{(n)} = 0, \quad \phi_n = \pi/2,  \quad \alpha_{n,m} = \beta_{n,m} =0, \nonumber\\
&& \mathrsfs{K}_{n,m} = 2\sqrt{U_p^{(n)}U_p^{(m)}}\cos\theta_{n,m}/(\Omega_m-\Omega_n), \nonumber\\
&& \mathrsfs{L}_{n,m} = 2\sqrt{U_p^{(n)}U_p^{(m)}}\cos\theta_{n,m}/(\Omega_m+\Omega_n).
\end{eqnarray}
where $U_p^{(n)} = ( E_1^{(n)} /2\Omega_n)^2$ are the pondermotive potentials and 
$\theta_{n,m}$ are angles between the fields $\E_1^{(n)}$ and $\E_1^{(m)}$, i.e.,  
$\cos\theta_{n,m} = \e_1^{(n)}\cdot\e_1^{(m)}$.

Equations (\ref{VolkovFact}), (\ref{FourierPhi1}), and (\ref{FourierPhi2}) form our 
general formalism. We point out that this formalism is not valid if one of the fields 
is a dc field (the generalization of the scheme to such a case will be published elsewhere). 
Having reached Eqs. (\ref{VolkovFact}), (\ref{FourierPhi1}), and (\ref{FourierPhi2}), the 
general form of multicolor ionization amplitudes for an arbitrary $N$ can be 
obtained within the SCVA as well as the SFA.  In the next section, we shall study ionization by a two-color 
laser field  ($N=2$), which is a most interesting case from the point of view of applications. 

\section{Ionization by a Bichromatic Laser Field}\label{Sec3}
\subsection{Two elliptically polarized fields}

The velocity gauge photoionization amplitudes in the case of a two-color laser field within the SFA and SCVA are given by 
\begin{widetext}
\begin{eqnarray}\label{Ampl_TwoColorElliptic}
\M_{\SFA, \,  \SCV} (\P)  =   -\pi i \sum_{n,m=-\infty}^{\infty} (-i)^{n+m} e^{-in\varphi_1 -im\varphi_2}\M_{\SFA, \, \SCV}^{(n,m)}  \delta\left( \frac{\P^2}2 + I_p + \left[2\mathrsfs{M}_1^{(1)}-n\right]\Omega_1 +\left[2\mathrsfs{M}_1^{(2)}-m\right]\Omega_2\right), 
\end{eqnarray}
where
\begin{eqnarray}
\M_{\SFA}^{(n,m)} &=& 2 \langle \P \ket{g} \left( \left[2\mathrsfs{M}_1^{(1)}-n\right]\Omega_1 +\left[2\mathrsfs{M}_1^{(2)}-m\right]\Omega_2\right)S_{n,m}, \\
\M_{\SCV}^{(n,m)} &=&  A_0^{(1)}\left[ M_{ph}^{(1)}(\pi-\phi_1)S_{n+1,m} + M_{ph}^{(1)}(\phi_1)S_{n-1,m}\right] 
+ A_0^{(2)}\left[M_{ph}^{(2)}(\pi-\phi_2)S_{n,m+1} + M_{ph}^{(2)}(\phi_2)S_{n,m-1} \right], \\ 
M_{ph}^{(n)}(\phi) &=& \bra{\psi_{\P}} \left(i\e_{1}^{(n)}\sin \phi +\e_{2}^{(n)}\cos \phi \right)\cdot \nabla \ket{g},\label{Ph-amplitude}\\
S_{n,m} &=& \sum_{k,l=-\infty}^{\infty} (-1)^l J_k(\mathrsfs{K}_{1,2}) J_l(\mathrsfs{L}_{1,2})
 \J_{n-l-k}\left(-\mathrsfs{N}_1, -\mathrsfs{M}_2^{(1)}; -2\varphi_0^{(1)} \right)  \nonumber\\
&& \times \J_{m-l+k}\left( -\mathrsfs{N}_2, -\mathrsfs{M}_2^{(2)}; -2\varphi_0^{(2)}\right)
 \exp\left\{ i\left( [n-l-k]\varphi_0^{(1)} + [m-l+k]\varphi_0^{(2)} + k\alpha_{1,2} + l\beta_{1,2} \right)\right\}. \label{DeffS} 
\end{eqnarray}
Here $\langle \P \ket{g}$ is the wave function of the initial state  in the momentum representation.
Note that neither $\M_{\SFA}^{(n,m)}$ nor $\M_{\SCV}^{(n,m)}$ depends on the absolute phases $\varphi_1$ and $\varphi_2$ of the laser fields.

Hereafter, we shall assume that $\Omega_2 > \Omega_1$ without lost of generality. Calculating the probability of ionization from Eq. (\ref{Ampl_TwoColorElliptic}), we discern that the result is determined whether $\Omega_1$ is commensurate with $\Omega_2$:

(i) {\it When $\Omega_1$ and $\Omega_2$ are noncommensurate}, the differential photoionization rate, which is the transition probability per unit time, reads 
\begin{eqnarray}\label{W_TwoColorElliptic_Noncomm}
W_{\SFA,\, \SCV}^{\rm (i)}(\P) = \sum_{n,m = -\infty}^{\infty}  \frac{\pi}2
\delta\left( \frac{\P^2}2 + I_p + \left[2\mathrsfs{M}_1^{(1)}-n\right]\Omega_1 +\left[2\mathrsfs{M}_1^{(2)}-m\right]\Omega_2\right)
 \left| \M_{\SFA,\, \SCV}^{(n,m)} \right|^2.
\end{eqnarray}
\end{widetext}
From Eq. (\ref{W_TwoColorElliptic_Noncomm}) one notices that the rate does not depend on the absolute phases of the laser fields.

(ii) {\it When $\Omega_1$ and $\Omega_2$ are commensurate}, we shall introduce the following notations $\Omega_2/\Omega_1 \equiv K_2/K_1$ and $\omega = \Omega_1/K_1\equiv\Omega_2/K_2$, where $K_1$ and $K_2$ are coprime (their greatest common divisor equals one). The choice of $\omega$ is unique, and we will discus its physical interpretation later. 

As an intermediate step in the derivation, we need to solve the following equation 
\begin{eqnarray}\label{DiophanEnergyConserv}
n = K_1 Q(n) + K_2 P(n),
\end{eqnarray}
where $n$ is a given integer, $Q(n)$ and $P(n)$ are unknown integers. Such an equation is called B\'{e}zout's identity \cite{Tignol2001}, which is a linear Diophantine equation \cite{Mordell1969}. Primarily, we summarize necessary evidences from the theory of Eq. (\ref{DiophanEnergyConserv}). Since $K_1$ and $K_2$ are coprime, a solution $[Q(n), P(n)]$ of this equation  always
exists and can be found by, e.g., the extended Euclid algorithm. Moreover, Eq. (\ref{DiophanEnergyConserv}) has infinity many solutions because if a pair  $[Q(n), P(n)]$ is a solution of Eq. (\ref{DiophanEnergyConserv}), then pairs $[Q(n)-mK_2, P(n)+mK_1]$, where $m$ being an arbitrary integer, are also solutions.

It is noteworthy to clarify the physical origin of Eq. (\ref{DiophanEnergyConserv}).  This equation follows from the law of conservation of energy, where $n$ is a number of quanta (the energy of each quantum equals $\omega$) absorbed by the electron and $Q(n)$ and $P(n)$ are numbers of quanta gained from the first and second laser fields, correspondingly.

Finally, we obtain the following expression for the differential photoionization rate:
\begin{widetext}
\begin{eqnarray}\label{W_TwoColorElliptic_Comm}
W_{\SFA,\, \SCV}^{\rm (ii)}(\P) &=& \sum_{n=N_0(\omega)}^{\infty} \frac{\pi}2 
\delta\left( \frac{\P^2}2 + I_p + 2\omega \left[K_1\mathrsfs{M}_1^{(1)}+K_2\mathrsfs{M}_1^{(2)}\right]-n\omega\right) \nonumber\\
&&\qquad \times\left| \sum_{m=-\infty}^{\infty} i^{m(K_2-K_1)} e^{im(K_2\varphi_1 - K_1\varphi_2)} 
\M_{\SFA,\, \SCV}^{(Q(n)-K_2m,\, P(n)+K_1 m)}  \right|^2,
\end{eqnarray}
where $N_0(\omega)$ is the minimal number of photons of the frequency $\omega$ absorbed. For the sake of convenience, we provide the differential photoionization rate for the special case of Eq. (\ref{W_TwoColorElliptic_Comm}) when the ratio $\Omega_2/\Omega_1\equiv K$ is an integer:
\begin{eqnarray}\label{W_TwoColorElliptic_CaseII}
 W_{\SFA, \, \SCV}^{\rm (iii)}(\P) = \sum_{n=N_0(\Omega_1)}^{\infty} \frac{\pi}2 
\delta\left( \frac{\P^2}2 + I_p + 2\Omega_1 \left[\mathrsfs{M}_1^{(1)}+K\mathrsfs{M}_1^{(2)}\right]-n\Omega_1\right)
\left| \sum_{m=-\infty}^{\infty} i^{m(K-1)} e^{im(K\varphi_1-\varphi_2)} \M_{\SFA,\, \SCV}^{(n-Km, m)}\right|^2,
\end{eqnarray}

%%%%%%%%%%%%%% BEGIN FIGURES %%%%%%%%%%%%
\begin{center}

\begin{figure}
\includegraphics[width=\columnwidth]{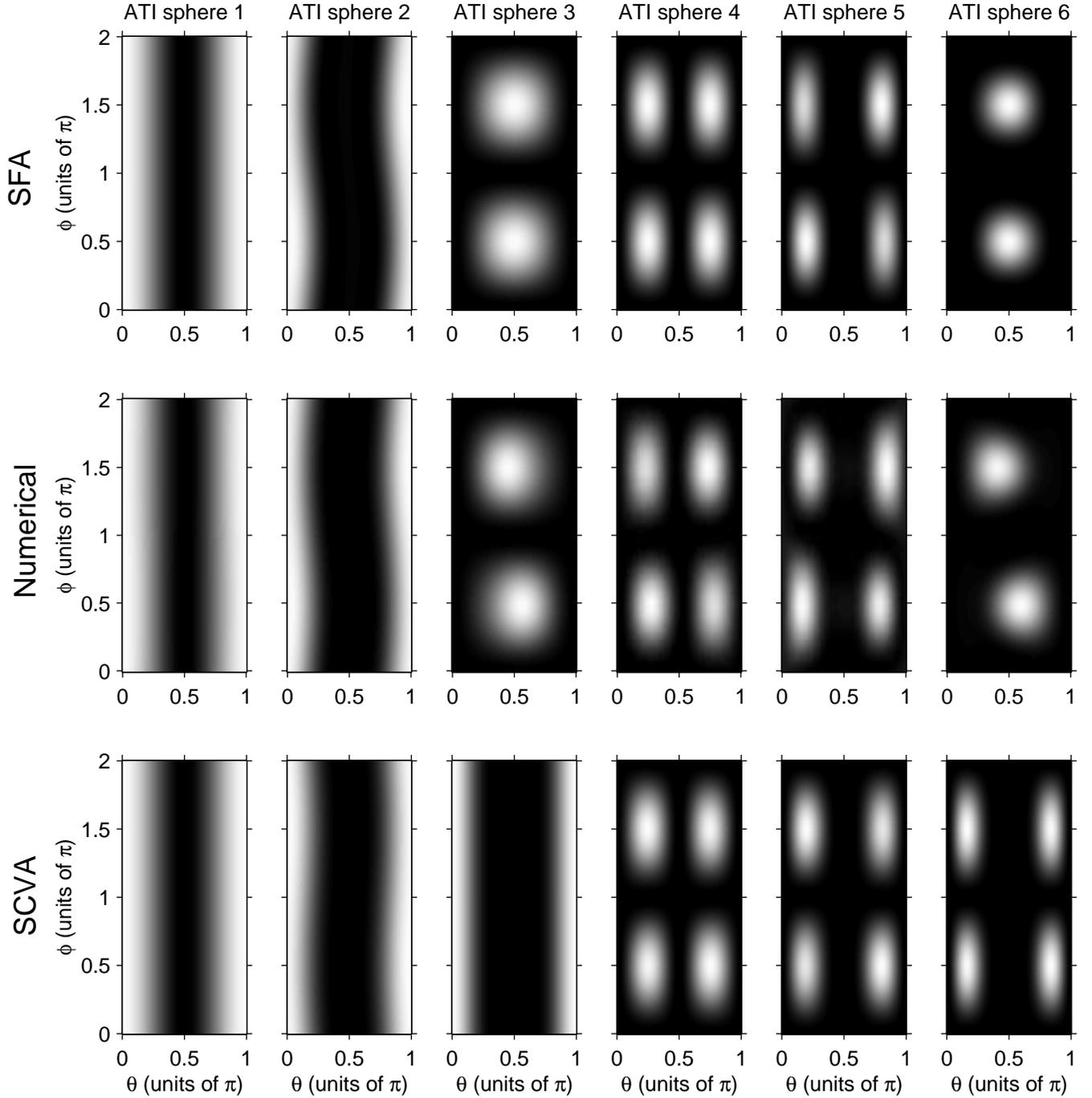}
\caption{Normalized ATI spheres (within the SFA, {\it ab initio}, and SCVA results) for the ground state of a hydrogen atom with field parameters given by $\E_1^{(1)}=(0,0,0.1)$ (a.u.), $\E_2^{(1)}={\bf 0}$, $\Omega_1=1$ (a.u), $\varphi_1 = 0$ and $\E_1^{(2)}=(0,0.1,0)$ (a.u.), $\E_2^{(2)}={\bf 0}$, $\Omega_2=3$ (a.u.), $\varphi_2 = 0$;  $\theta$ and $\phi$ are spherical angles (zenith and azimuth). Linear color scale goes from zero (black) to maximum (white).}\label{Fig1}
\end{figure}

\begin{figure}
\includegraphics[width=\columnwidth]{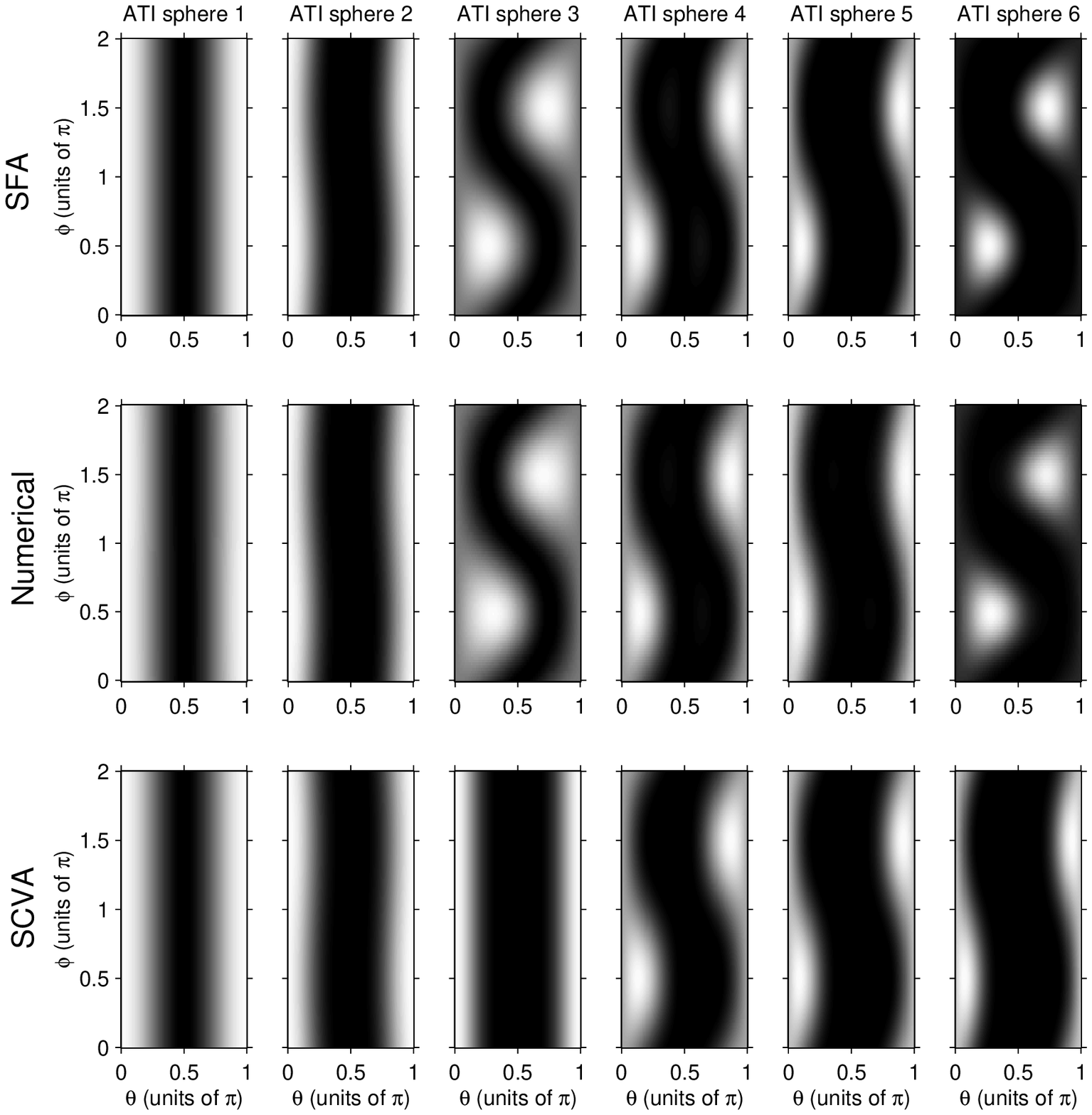}
\caption{Normalized ATI spheres (within the SFA, {\it ab initio}, and SCVA results) 
for the ground state of a hydrogen
atom with field parameters given by $\E_1^{(1)}=(0,0,0.1)$ (a.u.), $\E_2^{(1)}={\bf 0}$, $\Omega_1=1$ (a.u), $\varphi_1 = 0$ and
$\E_1^{(2)}=0.1\cdot(0, \sin[\pi/4], \cos[\pi/4])$ (a.u.), $\E_2^{(2)}={\bf 0}$, $\Omega_2=3$ (a.u.), $\varphi_2 = 0$. Linear color scale goes from zero (black) to maximum (white).}\label{Fig2}
\end{figure}

\end{center}
%%%%%%%% END FIGURES %%%%%%%%%%%%%
\end{widetext}
where $N_0(\Omega_1)$ is the minimal number of photons of the frequency $\Omega_1$ absorbed. 

Let us now interpret the parameter $\omega$. From the mathematical point of view, the set of 
arguments of the delta functions in Eq. (\ref{Ampl_TwoColorElliptic}) must coincide with the 
corresponding set of arguments in Eq. (\ref{W_TwoColorElliptic_Comm}). From the physical 
point of view, since stimulated processes of absorption of $n$ photons of the frequency 
$\Omega_2$ and subsequent emission of $m$ photons of $\Omega_1$ are possible in the presence 
of the two laser fields, then $\omega$ is the energy spacing between two neighboring ATI peaks.

If $K$ is large, then the sum over $m$ in Eq. (\ref{W_TwoColorElliptic_CaseII}) reduces to the term with $m=0$ (because $\M_{\SFA,\, \SCV}^{(n, m)}$ decreases exponentially when either of its indices is large), and therefore rate  (\ref{W_TwoColorElliptic_CaseII}) will not depend on the absolute phases. Similarly, rate (\ref{W_TwoColorElliptic_Comm}) eludes the dependence on the absolute phases in two cases: either $K_1$ or $K_2$ is large; $K_1$ and $K_2$ are simultaneously large. The physical origin of these statements is as follows: the effect of the absolute phases is to be manifested in ATI peaks that are induced by both the lasers simultaneously. Since $\Omega_2 K_1=\Omega_1 K_2$,  such a nearest peak corresponds to absorption of  $K_1$ quanta of $\Omega_2$ and $K_2$ quanta of $\Omega_1$. Therefore, the effect of the absolute phases is of $\max(K_1, K_2)$ order. 
 
\subsection{Two Linearly Polarized Fields: Special Cases}

In this section, we consider ionization by a bichromatic linearly polarized field. 

Computation of $S_{n,m}$ in Eq. (\ref{DeffS}) is itself a non-trivial problem. Nevertheless, Eq. (\ref{DeffS}) can be significantly reduced in the case when the coupling between two linearly polarized laser fields is small [this coupling is governed by $\Phi_{1,2}^{(2)}(t)$ in Eqs. (\ref{VolkovFact}) and (\ref{DeffPhi_One_Two})]. Introducing two dimensionless parameters
\begin{eqnarray}
\delta_{1,2} = \sqrt{U_p^{(1)} U_p^{(2)}}/\left(\Omega_2 \pm \Omega_1\right),
\end{eqnarray}
we define the {\it small coupling limit} by means of the following inequalities:    
\begin{eqnarray}\label{SmallCoupling}
\delta_1 \ll 1, \quad \delta_2 \ll 1.
\end{eqnarray}
Expanding the ordinary Bessel functions in Eq. (\ref{DeffS}) into a Taylor series with respect to small parameters $\mathrsfs{K}_{1,2}$ and $\mathrsfs{L}_{1,2}$, we obtain $S_{n,m}^L$ -- an asymptotic expansion of $S_{n,m}$ in the small coupling limit,
\begin{eqnarray}\label{SnmL}
S_{n,m}^L = J_n \left( -\mathrsfs{N}_1, -\frac{U_p^{(1)}}{2\Omega_1} \right)J_m\left(-\mathrsfs{N}_2, -\frac{U_p^{(2)}}{2\Omega_2} \right)  \nonumber\\
+  {\bf J}_1 {\bf \Delta}{\bf J}_2^T \cos\theta + O\left(\delta_1^2\right) + O(\delta_1\delta_2) + O\left(\delta_2^2\right),
\end{eqnarray} 
where $J_{n}(u, v) = \J_{n}(u,v; 0)$ is the two-dimensional Bessel function  (see, e.g., Refs. \cite{Reiss1980a, Dattoli1996, Reiss2003, Korsch2006}), $\theta\equiv\theta_{1,2}$ is the angle between the laser fields,  and the matrix ${\bf \Delta}$ and the vectors ${\bf J}_{1,2}$ are defined as 
\begin{eqnarray}
{\bf \Delta} &=& \left( \begin{array}{cc} \delta_2 & -\delta_1 \\ \delta_1 & -\delta_2 \end{array}\right), \\
{\bf J}_1 &=& \left[ J_{n-1} \left( -\mathrsfs{N}_1, -{U_p^{(1)}}/{2\Omega_1} \right), \right. \nonumber\\
&&\qquad \left. J_{n+1}\left(-\mathrsfs{N}_1, -{U_p^{(1)}}/{2\Omega_1} \right) \right], \\
{\bf J}_2 &=& \left[ J_{m+1} \left( -\mathrsfs{N}_2, -{U_p^{(2)}}/{2\Omega_2} \right), \right. \nonumber\\
&&\qquad \left. J_{m-1}\left(-\mathrsfs{N}_2, -{U_p^{(2)}}/{2\Omega_2} \right) \right].
\end{eqnarray}

Note that in the case when linearly polarized lasers are perpendicular ($\theta=\pi/2$), 
the full expression for $S_{n,m}$ [Eq.~(\ref{DeffS})] reduces immediately to
\begin{eqnarray}\label{SPerpendic}
S_{n,m} = J_n \left( -\mathrsfs{N}_1, -\frac{U_p^{(1)}}{2\Omega_1} \right)J_m\left(-\mathrsfs{N}_2, -\frac{U_p^{(2)}}{2\Omega_2} \right). 
\end{eqnarray} 
Equation (\ref{SPerpendic}) coincides with  the leading term in Eq. (\ref{SnmL}). 
Hence, the small coupling limit is an advantageous approximation only for nonperpendicular 
field configurations, i.e., $\theta\neq\pi/2$.

\subsection{Analytical Results v.s. Ab Initio Calculations}\label{SubSecTwoColor}

We shall consider the case of small couplings [see Eq. (\ref{SmallCoupling})] between two laser fields, and the ground state of a hydrogen atom shall be selected as the initial state. According to the delta function in Eq. (\ref{Ampl_TwoColorElliptic}), a photoelectron spectrum is confined to spheres. Therefore, to be able to completely analyze the momentum distribution of photoelectrons, we are to consider unfoldments of these ATI spheres. 
ATI spheres within the SFA, SCVA [Eq. (\ref{W_TwoColorElliptic_CaseII})], and {\it ab initio} calculations are
presented in Figs. \ref{Fig1} and \ref{Fig2}. 

The {\it ab initio} calculations were performed in spherical polar coordinates with the wave function expanded in
spherical harmonics $Y_l^m(\Omega)$, 
\begin{equation}
	\Psi(r,\Omega,t) = \sum_{l,\,m} R_{lm}(r,t) Y_l^m(\Omega)
\end{equation}
The resulting set of coupled one-dimensional radial Schr\"odinger equations including two laser fields,
\begin{eqnarray}
	 i\frac{\partial}{\partial t} R_{lm}(r,t) &=&  
	 \left[
	-\frac{1}{2} \nabla^2 +
	\frac {l(l+1)}{2r^2}
	+ V(r) \right] R_{lm}(r,t)
	 \nonumber \\ &
	-& \sum_{l',m'} \bra{Y_l^m} \E_1^{(1)}(t) \cdot\R |Y_{l'}^{m'}\rangle R_{l'm'}(r,t)
	 \nonumber \\ &
	-& \sum_{l',m'} \bra{Y_l^m} \E_1^{(2)}(t) \cdot \R |Y_{l'}^{m'}\rangle R_{l'm'}(r,t) \nonumber
\end{eqnarray}
is solved using finite-difference methods.
The angular laser coupling matrix elements are evaluated using 3j symbols.
The pulse shape was $f(t) = \exp[ -4 \ln2 (t/30)^2]$, where $\E_1^{(n)}(t) = f(t) \E_1^{(n)}\sin(\Omega_n t)$,
and the system was initially in the ground state.

The SFA within the velocity gauge agrees remarkably well with the results of numerical calculations;
nevertheless, the velocity gauge SCVA fails to agree. From {\it ab initio} as well as the SFA data, 
we readily notice that SCVA ATI spheres 3 and 6 are quite different from the corresponding 
SFA and numerical spheres.  Indeed, ATI sphere 3
should deviates substantially from ATI spheres 1 and 2 because it is the first sphere accessible 
to the $\Omega_2=3$ field, and further there may be imprinted a strong interference between
one-photon ionization by the field with $\Omega_2 = 3$ and three-photon ionization by the field with
$\Omega_1 =1$.  Similar arguments holds for ATI sphere 6.

Despite the similarities between the SFA and numerical results, small deviations remain.  The
numerical results exhibit a stronger left/right ($\phi=\pi/2$ and $\phi=3\pi/2$) symmetry breaking than 
seen in the SFA results, although the SFA does show some symmetry breaking, most notably in spheres 2 and 5.
Part of the symmetry breaking is captured by the SFA.  The presence of stronger symmetry breaking in the
numerical results may be related to Coulomb asymmetry effects previously seen in single-color elliptical
strong field ionization \cite{Goreslavski}.  In the present scenario, two fields of different frequencies
polarized at a relative angle generate ellipticity which could modify symmetries in the ATI spectra following
Ref.\cite{Goreslavski}. 

\section{Ionization by a Monochromatic Field}\label{Sec4}

In the previous section, it is seen that the SCVA inadequately describes ionization by a bichromatic laser field; moreover, the SFA has turned out to be a quite reliable approximation for the problem at hand.  We now consider the single-color scenario to see previous unnoticed deficiencies of the SCVA appear in this case  \cite{Yudin2007, Yudin2008}.

The differential rate of photoionization by a linearly polarized monochromatic laser field,
$$
\A(t) = -\E \sin(\Omega t)/\Omega,
$$
within the velocity gauge SCVA reads
\begin{eqnarray}
W_{\SCV}^{(L)}(\P) &=& \sum_{n=-\infty}^{\infty}\delta \left( \frac{\P^2}2 + I_p + U_p-n\Omega \right) \times  \nonumber\\
 &&2\pi U_p |M_{ph}(\pi/2 )|^2 \times \nonumber\\
	&& \left[ J_{n+1}(u_{0},v_{0}) + J_{n-1}(u_{0},v_{0}) \right]^2, \label{W_L}
\end{eqnarray}
where $U_p = (E/2\Omega)^2$ is the pondermotive potential, $u_{0}=-\P\cdot\E/\Omega^2$, $v_{0}=-U_p/2\Omega$.
For the sake of completeness, we quote the SFA differential photoionization rate for the linearly polarized laser field (the Keldysh-Faisal-Reiss theory \cite{Keldysh1965a, Faisal1973, Reiss1980a}, see also Ref. \cite{Becker2005a}) written in a suitable language for our discussion,  
\begin{eqnarray}
W_{\SFA}^{(L)} (\P) &=& \sum_{n=-\infty}^{\infty} \delta\left( \frac{\P^2}2 + I_p+ U_p - n\Omega \right)\times \nonumber\\
 && 2\pi (U_p - n\Omega)^2 \left| \bra{\P} g\rangle J_n \left(u_0, v_0\right) \right|^2. \label{W_SFA}
\end{eqnarray}
These equations are known in the literature and can be derived using the same formalism present above by starting with a single-color form for the field. 

\begin{figure}
\begin{center}
\includegraphics[width=\columnwidth]{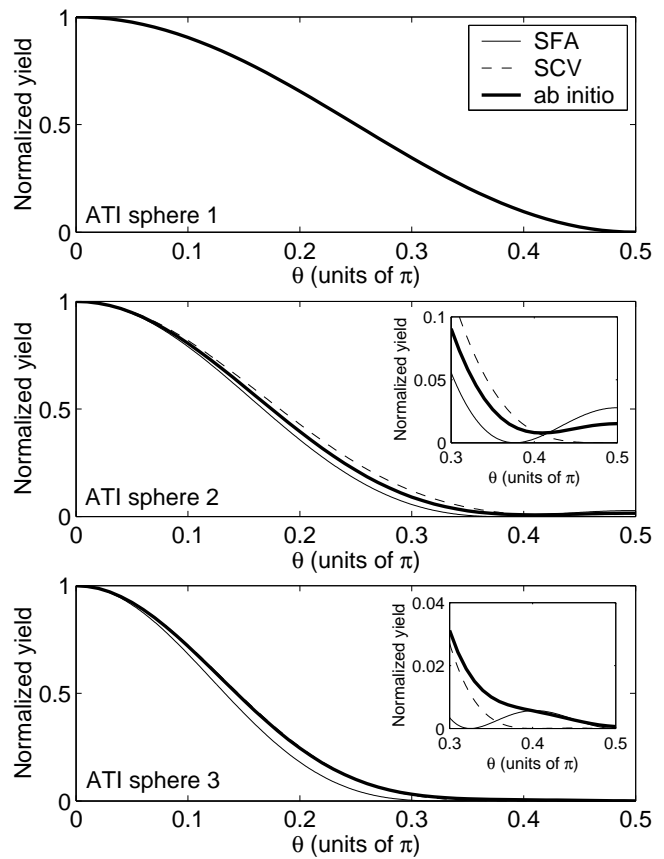}
\end{center}
\caption{Normalized ATI cuts within the SCVA [Eq.  (\ref{W_L})], SFA [Eq.  (\ref{W_SFA})], and {\it ab initio} results for the ground state of a hydrogen atom in the linearly polarized monochromatic laser field with $E = 0.1$ (a.u.) and $\Omega = 2$ (a.u.).  The figures correspond to the first ATI peak (top), 
second peak (middle), and third peak (bottom).  The thin solid lines are the SFA results, 
dashed lines are the SCVA results, and  the thick solid lines are the {\it ab initio} results. 
Note that all lines overlap in the top plot. 
The insets show magnifications of the corresponding plots in the vicinity of $\theta=\pi/2$. }
\label{Fig3}
\end{figure}

Let us illustrate and test the analytical formulas. In the case of the monochromatic field, 
ATI spheres are cylindrically symmetric for the ground state of a hydrogen atom, viz., 
momentum distributions do not depend on the azimuthal angle $\varphi$; thus, 
we present cuts along the zenith angle $\theta$. Such cuts are presented in Fig. \ref{Fig3}. From the figure, one concludes that the SFA and SCVA are not so radically different as in the 
bichromatic case. Note that the  normalized first ATI peaks within the SCVA, SFA, and {\it ab initio} 
treatment are the same (top panel of Fig.~\ref{Fig3}).  However, for the second ATI peak, 
the velocity gauge SFA slightly better reproduces the results of numerical simulations in the vicinity 
$\theta = \pi/2$ than the velocity gauge SCVA. The SCVA momentum distributions, 
given by Eq. (\ref{W_L}), always equal to zero at $\theta=\pi/2$ because of the presence 
of the one-photon ionization amplitude $M_{ph}$ whereas the {\it ab initio} and SFA both give 
nonz-ero yield at $\theta = \pi/2$.  Such nonzero yield at $\theta = \pi/2$ appears for all even
order peak in monochromatic ionization (not shown), and in each case SCVA fails to capture this yield 
while SFA at least gives a nonzero yield.

\section{Conclusions and Discussions}

The general analytical formalism for calculating SFA and SCVA multicolor ionization amplitudes are presented. Analyzing photoelectron spectra induced by a bichromatic laser field in the regime when smallest frequency of the laser field is larger than the ionization
potential of the atom, we conclude that the velocity gauge SFA is more adequate than the velocity gauge SCVA. Note that the length gauge SFA better describes photoelectron momentum distributions than the velocity gauge SFA in the case of ionization by a  low frequency field (see, e.g., Ref. \cite{Bauer2008}). As the next step, one may try to compare the length gauge SCVA and SFA; there is no simple analytical approach to the problem within the length gauge. However, we suppose that the length gauge might not lead to qualitatively different results since the canonical momentum $\P$ approaches the kinetic momentum $\K=\P + \A(t)$ in the high frequency limit. 

The origin of the observed qualitative difference between the velocity gauge SFA and SCVA is related to the term $\A^2(t)/2$ in the definition of the velocity gauge SFA ionization amplitude [Eq. (\ref{Deff_M_SFA})]. If we had ignored that term in Eq. (\ref{Deff_M_SFA}), we would have obtained qualitatively the same results as by the SCVA. Ostensibly, the minor difference between the SFA and SCVA for the case of ionization by monochromatic radiation (Sec. \ref{Sec4}) is enhanced through interference between different one-photon ionization channels in the presence of a bichromatic laser field (Sec. \ref{SubSecTwoColor}).

Finally, we point out that the formalism developed in Sec. \ref{Sec2} may have two further applications in addition to its original purpose. 

The first one is two-electron strong field phenomena in the presence of a multicolor laser field. As it has been shown in Refs. \cite{Bergou1981, Faisal1994a}, there exists the exact solution of the Schr{\"o}dinger equation for two interacting electrons in a laser field. Moreover, this solution is a product of ``Volkov-type'' phase (\ref{TimeDepVolkov})  and a time-independent coordinate part. (Regarding some applications of the exact solution \cite{Faisal1994a} to two-electron phenomena induced by a single-color laser field see review \cite{Becker2005a}.)

The second application is to study analytically the effects of a pulse envelope, which are of ongoing interest. In particularly, the effect of a carrier-envelope phase on ionized electron momentum can be investigated (see, e.g., Ref. \cite{Peng2008}): the vector potential used in Eqs. (5) and (6) in Ref. \cite{Peng2008} is equivalent to a three-color case. In this formulation, the carrier-envelope phase effects will be the effects of the absolute phases.

\acknowledgments

We would like to thank M. Yu. Ivanov and S. Patchkovskii for valuable discussions. Financial support to D.I.B. by an NSERC SRO grant is gratefully acknowledged.

\bibliography{MultiColor}
\end{document}